\documentclass[12pt]{article}
\usepackage{amstex}

\def\Htilde{\tilde{H}}
\def\Ytilde{\tilde{Y}}
\def\Atilde{\tilde{A}}
\def\ep{\epsilon}
\def\Pf{{\rm Pf}}
\def\tr{{\rm tr}}

\begin{document}
\vspace*{-.6in}
\thispagestyle{empty}
\begin{flushright}
CALT-68-2103\\
hep-th/9703062
\end{flushright}
\baselineskip = 20pt

\vspace{.5in}
{\Large
\begin{center}
Wrapping the M Theory Five-Brane on K3\end{center}}

\begin{center}
Sergey Cherkis and John H. Schwarz\footnote{Work
supported in part by the U.S. Dept. of Energy under Grant No.
DE-FG03-92-ER40701.}
\\
\emph{California Institute of Technology, Pasadena, CA  91125, USA}
\end{center}
\vspace{1in}

\begin{center}
\textbf{Abstract}
\end{center}
\begin{quotation}
\noindent  Using a recently constructed M5-brane 
world-volume action, we demonstrate that wrapping the M5-brane on 
K3 gives the heterotic string in seven dimensions. To facilitate the comparison, 
a new version of the world-sheet action for the Narain-compactified
heterotic string, with manifest T duality invariance, is formulated.
\end{quotation}
\vfil

\newpage

\pagenumbering{arabic}

\section{Introduction}

One of the remarkable dualities that was pointed out
by Witten two years ago relates heterotic string theory
compactified on a torus to seven dimensions and
eleven-dimensional M theory (a name that was coined later)
compactified on K3~\cite{witten}. It was noted that the Narain moduli
space of the 7d heterotic theory precisely matches the moduli space of 
Ricci-flat metrics on K3. Points of enhanced gauge symmetry
in the heterotic description correspond to K3's in which
two-cycles degenerate. Also, the correspondence of the
volume modulus ${\cal V}$ of the
K3 and the heterotic dilaton $\phi$ satisfies the scaling
law ${\cal V}\sim \lambda_H^{4/3}$, where $\lambda_H = {\rm exp}\, \phi$
is the heterotic string coupling constant. Thus, this pair of 
theories is U dual in the sense that large volume corresponds
to strong coupling.

Shortly after this duality was proposed, the matching of various
solitons in 7d, viewed from the two dual perspectives, was 
considered~\cite{harvey,townsend}.\footnote{together with the related 6d problem
of matching solitons of type IIA theory on K3 and heterotic theory on T${}^4$.}
It was noted, in particular,
that the heterotic string in 7d can be understood from the
M theory viewpoint as arising from wrapping the five-brane
of M theory (or the `M5-brane') on the K3. A convincing
case was made that this give the correct degrees of freedom on
the string world sheet. Specifically, the 3 right-moving and
19 left-moving compact chiral bosons on the world sheet arise
from the zero modes of the chiral two-form of the M5-brane
as a direct consequence of the fact that K3 has 3 self-dual
and 19 anti-self-dual two-forms. The purpose of this paper is to
study this correspondence in more detail. This turns out to
be an interesting, and possibly useful, exercise. In particular,
it motivates the formulation of a new world-sheet action for the
Narain-compactified heterotic string in which the T duality symmetries
are manifest.

Explicit formulas for the M5-brane world-volume
action have recently been constructed~\cite{perry,pasti1,bandos,aganagic},
making the present study possible. Ref.~\cite{perry} 
developed a formalism without manifest 
covariance for describing the chiral two-form gauge field of the M5-brane.
A covariant formalism was subsequently applied to this problem by Pasti,
Sorokin, and Tonin (PST)~\cite{pasti1} using an approach they
had developed earlier~\cite{pasti2}. This formulation contains an
auxiliary scalar field and new gauge invariances.\footnote{An alternative
covariant formulation of the M5-brane field equations, which
does not appear to contain this scalar field, has been proposed in ref.~\cite{howe}.} 
The non-covariant
formulas of~\cite{perry} arise in the PST formulation from choosing
a particular gauge. This paper will use the covariant PST approach.
When four of the dimensions of the M5-brane are wrapped on a K3
space, and only zero modes of the K3 are retained (as in Kaluza-Klein
dimensional reduction), the remaining action describes the two-dimensional
world-sheet of a string in seven dimensions. Since the resulting formulas
look quite different from anything that can be found in the literature,
the first part of this paper is devoted to developing the appropriate description of
the heterotic string theory. Then it is relatively straightforward to see
that this corresponds to what arises 
from the double dimensional reduction of the M5-brane
on K3. Most of the essential issues already arise for the bosonic degrees of 
freedom of the M5-brane and the heterotic string. Therefore, in this paper we
only consider this truncated problem. It would be desirable, however, 
to generalize our analysis to include the fermionic degree of freedom
so as to incorporate the appropriate target space supersymmetries and
world-sheet kappa symmetry.

\medskip
\section{The Heterotic String}

There are two main approaches to constructing the world-sheet action of the
heterotic string that have been used in the past~\cite{gross}.  
In one of them, the internal
torus is described in terms of bosonic coordinates. The fact that these
bosons are chiral ({\it i.e.}, the left-movers and right-movers behave differently)
is imposed through external constraints.  In the second approach these bosonic
coordinates are replaced by world-sheet fermions, which are Majorana--Weyl in
the 2d sense.  What will be most convenient for our purposes is a variant of
the first approach.  In this variant the coordinates of the Narain torus are still
represented by bosonic fields, but the chirality of these fields is achieved
through new gauge invariances rather than external constraints.

In order not to confront too many issues at once, the
lagrangian of the Narain torus for a flat world-sheet geometry
will be described first.  Then we will
add a world-sheet metric field so as to achieve general coordinate invariance
of the world-sheet theory.  Finally,  the auxiliary
world-sheet metric will be eliminated
by solving its equations of motion, so as to obtain the
appropriate generalization of the Nambu--Goto action.

\subsection{Lorentz Invariant Action for the Narain Torus}

Consider the Narain compactified heterotic string in a
Minkowski space-time with $d = 10 - n$ dimensions.  
Let these coordinates be denoted
by $X^m$ with $m = 0,1, \ldots, d - 1 = 9 - n$.  To properly account for all
the degrees of freedom, the Narain torus should be described by $16 + 2n$
bosonic coordinates $Y^I, $ $I = 1,2, \ldots, 16 + 2n$.  These will be arranged to
describe $26 - d = 16 + n$ left-movers and $10 - d = n$ right-movers.  The
$Y^I$ are taken to be angular coordinates, with period $2\pi$, so that $Y^I
\sim Y^I + 2\pi$, and the conjugate momenta are integers.  The actual size and
shape of the torus is encoded in a matrix of moduli, denoted $M_{IJ}$, which
will be described below.

The $(16 + 2n)$-dimensional
lattice of allowed momenta should form an even self-dual lattice of signature
$(n, 16 + n)$.  Let us therefore introduce a matrix
\begin{equation}
\eta = \left(\begin{matrix}
I_n & 0\\
0 & - I_{16+n}\end{matrix}\right) ,
\end{equation}
where $I_n$ is the $n \times n$ unit matrix.  An
even self-dual lattice with this signature has a set of $16 + 2n$ basis
vectors $V_I$, and the symmetric matrix
\begin{equation}
L_{IJ} = V_I^a \eta_{ab} V_J^b
\end{equation}
characterizes the lattice.  A convenient specific choice is
\begin{equation}
L = \Lambda_8 \oplus \Lambda_8 \oplus \sigma \oplus \ldots \oplus \sigma,
\end{equation}
where $\Lambda_8$ is the negative of the $E_8$ Cartan matrix and $\sigma =
\left(\begin{matrix} 0 & 1\\ 1 & 0 \end{matrix}\right)$ appears $n$ times.  For
$n > 0$, an arbitrary even self-dual lattice can be obtained from this one by
applying a suitable $O (n, 16 +n)$ transformation.  For $n = 0$, this choice
is appropriate to the $E_8 \times E_8$ theory.  For the $SO(32)$ theory in 10d
one would use $L = \Lambda_{16}$, which characterizes the weight lattice of
the group spin(32)/${\bf Z}_2$.

The Narain moduli space is characterized, up to $T$ duality equivalences that
will be discussed below, by a symmetric matrix $M'_{ab} \in O (n, 16 +
n)$, which satisfies $M'\eta M' = \eta$. 
The fact that it is symmetric means that it actually parametrizes the
coset space $O(n, 16 + n)/O(n) \times O (16+n)$, which has $n(16 + n)$ real
dimensions.  To describe the $T$ duality equivalences, it is convenient to
change to the basis defined by the basis vectors of the self-dual lattice.
Accordingly, we define
\begin{equation}
M_{IJ} = V_I^a M'_{ab} V_J^b = (V^T M' V)_{IJ}.
\end{equation}
This matrix is also symmetric and satisfies
\begin{equation}
ML^{-1} M = L,
\end{equation}
from which it follows that $(L^{-1}M)^2 = 1$.  This allows us to define
projection operators
\begin{equation}
{\cal P}_\pm = {1\over 2} (1 \pm L^{-1} M).
\end{equation}
${\cal P}_+$ projects onto an $n$-dimensional subspace, which will correspond
to right-movers.  Similarly, ${\cal P}_-$ projects onto the
$(16+n)$-dimensional space of left-movers.

The theory we are seeking should be 
invariant under an infinite discrete group of $T$ duality
transformations, denoted 
$\Gamma_{n,16+n}$,\footnote{It is often called $O(n,16 + n; {\bf Z})$.} 
so that the actual moduli space is the standard Narain space
\begin{equation}
{\cal M}_{n, 16 + n} = \Gamma_{n, 16 + n}\backslash O(n, 16 + n) / O(n) \times O (16 +n).
\end{equation}
The lagrangian will be formally invariant under an $O(n, 16 + n)$
transformation.  In terms of coordinates $Y' = VY$, the transformation is just
$M' \rightarrow \Lambda^T M' \Lambda$ and $ Y' \rightarrow \Lambda Y'$, where $
\Lambda \in O (n, 16 + n)$.  In terms of unprimed quantities, this is  $Y
\rightarrow \tilde{\Lambda} Y$ and $ M \rightarrow \tilde{\Lambda}^T M
\tilde{\Lambda}$, where $\tilde{\Lambda} = V^{-1} \Lambda V$.  This is actually
only a true symmetry of the theory when $Y \rightarrow \tilde{\Lambda} Y$
respects the periodicities of the $Y$'s.  In other words, the $T$ duality
transformations form a discrete group $\Gamma_{n, 16 + n}$ for which all the
entries of the matrix $\tilde{\Lambda}$ are integers.

The desired equations of motion for the $Y$ 
coordinates are~\cite{cecotti,duff,tseytlin,maharana}
\begin{equation}
{\cal P}_- \partial_+ Y = 0 \quad { \rm and} \quad {\cal P}_+ \partial_- Y = 0,
\label{Yeqs}
\end{equation}
where $\xi^\pm = \xi^1 \pm \xi^0$, so that $\partial_\pm = {1\over 2}
(\partial_1 \pm \partial_0)$.  $\xi^0$ and $\xi^1$ are the world-sheet time and
space, respectively. The pair of equations in (\ref{Yeqs}) can be combined
in the form
\begin{equation}
M\partial_0 Y - L\partial_1 Y = 0.
\end{equation}
It is easy to write down a lagrangian that gives this equation:
\begin{equation}
{\cal L}_N = {1\over 2} (\partial_0 Y M \partial_0 Y - \partial_0 Y L \partial_1 Y).
\end{equation}
This is just (a modest generalization of) the lagrangian
for a chiral 2d boson first proposed by Floreanini and 
Jackiw~\cite{floreanini}.  
Two things are peculiar about this lagrangian.  First, it does not
have manifest Lorentz invariance.  
However, in ref.~\cite{schwarz} it
was shown that ${\cal L}_N$ has a global symmetry that can be interpreted as
describing a non-manifest Lorentz invariance. 
Second, it gives the equation of motion
\begin{equation}
\partial_0 [M\partial_0 Y - L\partial_1 Y] = 0,
\end{equation}
which has a second, unwanted, solution $Y^I = f^I (\xi^1)$.  The resolution of
the second problem is quite simple.  The transformation $\delta Y^I = f^I
(\xi^1)$ is a gauge symmetry of ${\cal L}_N$, and therefore $f^I (\xi^1)$ represents
unphysical gauge degrees of freedom.

The first problem, the noncovariance of ${\cal L}_N$, is more interesting.  
This paper will follow the PST approach~\cite{pasti2}, 
and extend ${\cal L}_N$ to a manifestly Lorentz
invariant action by introducing an auxiliary scalar field $a(\xi)$.  The
desired generalization of ${\cal L}_N$ is then
\begin{equation}
{\cal L}_{PST} = {1\over 2(\partial a)^2} (\tilde{Y} M \tilde{Y} - \tilde{Y} L\,
\partial Y \cdot \partial a),
\end{equation}
where
\begin{equation}
\tilde{Y}^I = \epsilon^{\alpha\beta} \partial_\alpha Y^I \partial_\beta a.
\label{Ytilde}
\end{equation}
Also, $(\partial a)^2$ and $ \partial Y \cdot \partial a$ are formed using the 2d
Lorentz metric, which  is diagonal with $\eta^{00} = - 1$ and $\eta^{11} = 1$.  

The theory given by ${\cal L}_{PST}$ has two gauge invariances.  The first is
\begin{eqnarray}
\delta Y &=& \varphi \left({1\over \partial_+ a} {\cal P}_- \partial_+ Y +
{1\over\partial_- a} {\cal P}_+ \partial_- Y\right),\nonumber \\
\delta a &=& \varphi,
\end{eqnarray}
where $\varphi (\xi^0, \xi^1)$ is an arbitrary infinitesimal scalar function.
If this gauge freedom is used to set $a = \xi^1$, then ${\cal L}_{PST}$ reduces to
${\cal L}_N$.  The second gauge invariance is
\begin{equation}
\delta Y^I = f^I (a), \quad \delta a = 0, \label{fifteen}
\end{equation}
where $f^I$ are arbitrary infinitesimal functions of one variable.  This is the
covariant version of the gauge symmetry of ${\cal L}_N$ that was used to argue that the
undesired solution of the equations of motion is pure gauge.

\subsection{Reparametrization Invariant Action}

The formulas described above are not the whole story of the bosonic degrees of
freedom of the toroidally compactified heterotic string, because they lack the
Virasoro constraint conditions.  The standard way to remedy this situation is
to include an auxiliary world-sheet metric field 
$g_{\alpha\beta}(\xi)$, so that the world-sheet
Lorentz invariance is replaced by world-sheet general coordinate invariance.
Since we now want to include the coordinates $X^m$ describing the uncompactified
dimensions, as well, let us also introduce an induced world-sheet metric
\begin{equation}
G_{\alpha\beta} = g_{mn}(X) \partial_{\alpha}X^m\partial_{\beta} X^n,
\end{equation}
where $g_{mn}(X)$ is the string frame target-space metric in $d$ dimensions.
It is related to the canonically normalized metric by a factor of the form
${\rm exp} (\alpha\phi)$, where $\phi$ is the dilaton and $\alpha$ is a numerical
constant, which can be computed by requiring that the target-space lagrangian is
proportional ${\rm exp} (-2\phi)$.  We will mostly be
interested in taking $\phi$ to be a constant and $g_{mn}$ to be proportional
to the flat Minkowski metric. Then the heterotic string coupling constant
is $\lambda_H = {\rm exp}\, \phi$, and the desired world sheet lagrangian is
\begin{equation}
{\cal L}_g = - {1\over 2} \sqrt{-g} g^{\alpha\beta} G_{\alpha\beta}
+ {\tilde{Y} M \tilde{Y}\over 2\sqrt{-g} (\partial a)^2} - {\tilde{Y} L\,
\partial Y \cdot \partial a\over 2(\partial a)^2} . \label{Laux}
\end{equation}
Now, of course, $(\partial a)^2 = g^{\alpha\beta} \partial_\alpha a
\partial_\beta a$ and $ \partial Y \cdot \partial a = g^{\alpha\beta}
\partial_\alpha Y \partial_\beta a$.  The placement of the $\sqrt{-g}$ factors
reflects the fact that $\tilde{Y}/\sqrt{-g}$ transforms as a scalar. The complete
lagrangian should also contain a term of the form $\phi\sqrt{-g} R$, where $R$ is the
world-sheet curvature scalar. However, this term is an order $\alpha'$ correction,
which is beyond the scope of the present analysis.

There are a few points to be made about ${\cal L}_g$.  First of all, the PST gauge
symmetries continue to hold, so it describes the correct degrees of freedom.
Second, just as for more conventional string actions, it has Weyl invariance:
$g_{\alpha\beta} \rightarrow \lambda g_{\alpha\beta}$ is a local symmetry.
This ensures that the stress tensor
\begin{equation}
T_{\alpha\beta} = - {2\over\sqrt{-g}} {\delta S_g\over\delta g^{\alpha\beta}},
\end{equation}
is traceless $(g^{\alpha\beta} T_{\alpha\beta} = 0)$.

Using the general coordinate invariance to choose $g_{\alpha\beta}$ conformally
flat, and using the PST gauge invariance to set $a = \xi^1$, the $Y$ equations
of motion reduce to those described in the previous subsection.  In addition,
one obtains the classical Virasoro constraints $T_{++} = T_{--} = 0$.  With
these gauge choices one finds
\begin{equation}
T_{++} = G_{++} + \partial_0 Y (M \partial_0 Y -
L\partial_1 Y) + \partial_0 Y L \partial_+ Y.
\end{equation}
Imposing the equation of motion $M\partial_0 Y - L \partial_1 Y = 0$, this can
be recast in the form
\begin{equation}
T_{++} = G_{++} +  \partial_+ Y L \partial_+ Y,
\end{equation}
which is the standard result for the left-moving stress tensor.  In similar
fashion one finds
\begin{equation}
T_{--} = G_{--} +  \partial_- Y L \partial_- Y,
\end{equation}
for the right-movers.

\subsection{Elimination of the Auxiliary Metric}

The lagrangian ${\cal L}_g$ is written with an auxiliary world-volume metric, which is
called the Howe--Tucker or Polyakov formulation.  This is the most convenient
description for many purposes.  However, for the purpose of comparing to
expressions derived from the M5-brane later in this paper, it will be useful to
also know the version of the lagrangian in which the auxiliary metric is
eliminated --- the Nambu--Goto formulation.

Note that ${\cal L}_g$ only involves the metric components in the combination
$\sqrt{-g} g^{\alpha\beta}$.  Let us therefore define
\begin{equation}
p = \sqrt{-g} g^{11} \quad {\rm and} \quad  q = \sqrt{-g} g^{10}.
\end{equation}
Since $\det (\sqrt{-g} g^{\alpha\beta}) = - 1$, the remaining component is
\begin{equation}
\sqrt{-g} g^{00} = {q^2 - 1\over p}.
\end{equation}
With these definitions, and setting $a = \xi^1$ for simplicity, ${\cal L}_g$
in eq.~(\ref{Laux}) takes the form
\begin{equation}
{\cal L}_g = pA_1 + q A_2 + {q^2\over p} A_3 + {1\over p} A_4 + {q\over p} A_5 + A_6,
\end{equation}
where
\begin{eqnarray}
A_1 &=& - {1\over 2} G_{11}, \quad A_2 = - G_{01}, 
\quad A_3 = - {1\over 2} G_{00}\nonumber
\\
A_4 &=& {1\over 2} (G_{00} + \tilde{Y} M \tilde{Y})\nonumber \\
A_5 &=& - {1\over 2} \tilde{Y} L \, \partial_0 Y, \quad A_6 = - {1\over 2} \tilde{Y}
L\partial_1 Y.
\end{eqnarray}

It is straightforward to form and solve the $p$ and $q$ equations of motion.
Substituting the solutions back in ${\cal L}_g$ gives the lagrangian
\begin{equation}
{\cal L} = - \sqrt{4\left(A_1 - {A_2^2\over 4A_3}\right) \left(A_4 - {A_5^2\over
4A_3}\right)} + A_6 - {A_2 A_5\over 2A_3}.
\end{equation}
Substituting the expressions for the $A$'s, given above, and reinstating the
$a$ dependence, leaves
the final form for the bosonic part of the heterotic string in
$10 - n$ dimensions
\begin{equation}
{\cal L} = - \sqrt{-G}\sqrt{1 + {\tilde{Y} M \tilde{Y}\over G(\partial a)^2} +
\left({\tilde{Y} L \tilde{Y}\over 2G(\partial a)^2}\right)^2 }- {\tilde{Y}
L\partial Y \cdot \partial a\over 2(\partial a)^2}, \label{hetfinal}
\end{equation}
where $G = \det G_{\alpha\beta}$, and now
\begin{equation}
(\partial a)^2 = G^{\alpha\beta} \partial_\alpha a \partial_\beta a, \quad \partial Y
\cdot \partial a = G^{\alpha\beta} \partial_\alpha Y \partial_\beta a.
\end{equation}

\medskip
\section{Wrapping the M-Theory Five-Brane on K3}

Let us now consider double dimensional reduction of the M5-brane 
on K3.\footnote{See ref.~\cite{aspinwall} for a review of the mathematics of K3
and some of its appearances in string theory dualities.} 
Our starting point is the bosonic part of the M5-brane action~\cite{perry}
in the general coordinate invariant PST formulation of~\cite{pasti1}
\footnote{The supersymmetric extension including the fermionic
$\theta$ variables is described in refs.~\cite{bandos,aganagic}.}
\begin{equation}
\label{5brane}
{\cal L}_5 =-\sqrt{-\det\left( G_{\mu\nu}+i\frac{G_{\mu\rho}G_{\nu\sigma}
\Htilde^{\rho\sigma}}
{\sqrt{-({\rm det}\,G_{\mu\nu}) (\partial a)^2}}\right)} - \frac{\Htilde^{\mu\nu} H_{\mu\nu\rho} 
G^{\rho\lambda} \partial_\lambda a}{4 (\partial a)^2},
\end{equation}
where $G_{\mu\nu} = g_{MN}(X)\partial_{\mu}X^M
\partial_{\nu}X^N$ is the induced world-volume metric, 
$H_{\mu\nu\rho} = 3\partial_{[\mu} B_{\nu\rho]}$,
$(\partial a)^2=G^{\mu\nu}\partial_\mu a \partial_\nu a$, and
\begin{equation}
\Htilde^{\mu\nu}=\frac{1}{6} \epsilon^{\mu\nu\rho\sigma\tau\lambda} 
H_{\rho\sigma
\tau} \partial_\lambda a.
\end{equation}
Refs.~\cite{perry,aganagic} only considered flat backgrounds. However, the generalization
to include $g_{MN}(X)$ in the way indicated is certainly allowed, provided
that it solves the 11d supergravity field equations.  
Since the other 11d fields are still assumed to vanish, 
$g_{MN}(X)$ must be Ricci flat. We will take it to be a product of a Ricci-flat
K3 and a flat 7d Minkowski space-time.

The action $ \int {\cal L}_5 d^6 \sigma$ has several types of local symmetries.
The two manifest ones are 6d general coordinate invariance and invariance
under a gauge transformation $\delta B = d\Lambda$. In addition, there are
two PST gauge invariances~\cite{pasti1}. One, 
with parameters $\phi_{\mu}(\sigma)$, is given by\footnote{Actually, this symmetry
and $\delta B = d \Lambda$ can be combined and generalized to $\delta B = f(a) d \Lambda$,
where $f$ is an arbitrary function of $a$ and $\Lambda$ is an arbitrary one-form.
This is the two-form counterpart of eq.~(\ref{fifteen}).}
\begin{equation}
\delta B_{\mu\nu} = \phi_{\mu} \partial_{\nu} a - \phi_{\nu} \partial_{\mu} a,
\end{equation}
while $a$ and $X^M$ remain invariant. The other, 
with parameter $\varphi(\sigma)$, is given by $\delta a = \varphi$, $\delta X^M =0$,
and
\begin{equation}
\delta B_{\mu\nu} = {1 \over (\partial a)^2} \varphi H_{\mu\nu\rho}
G^{\rho\lambda} \partial_{\lambda} a - 2 \varphi {\partial L_1 \over
\partial \tilde H^{\mu\nu} },
\end{equation}
where $L_1$ is the first term in eq.~(\ref{5brane}).

Since the M5-brane is taken to wrap the spatial K3, 
the diffeomorphism invariances of
the M5-brane action in these dimensions can be used to equate 
the four world-volume  coordinates  that describe the K3 
with the four target-space coordinates that describe the K3. In other words,
we set $\sigma^{\mu} = ( \xi^{\alpha}, x^i)$ and $X^M = (X^m, x^i)$.
Note that Latin indices $i,j,k$ are used for the K3 dimensions $(x^i)$ 
and early Greek letters for the directions $(\xi^\alpha)$,
which are the world-sheet coordinates of the resulting string action.
This wrapping by identification of coordinates, together with the extraction of the
K3 zero modes, is what is meant by double dimensional reduction. 
With these choices, the 6d metric can be decomposed into blocks
\begin{equation}
\label{metricdecomposition}
\left(G_{\mu\nu}\right)=\left( \begin{array}{cc} 
                                    \tilde G_{\alpha\beta} & 0 \\ 
                                      0   &  h_{ij}    
                                          \end{array} \right) ,
\end{equation}
with $h_{ij}$ and $\tilde G_{\alpha\beta}$ being the K3 metric and the induced
metric on the string world-sheet, respectively. The purpose of the tilde is to
emphasize that $\tilde G_{\alpha\beta} = \tilde g_{mn} \partial_{\alpha} X^m
\partial_{\beta} X^n$, where $\tilde g_{mn}$ is the 7d part of the
canonical 11d metric. It differs from $g_{mn}$ of sect. 2 by a scale factor,
which will be determined below. It is
convenient to take the PST scalar field $a$
to depend on the $\xi^{\alpha}$ coordinates only. This amounts to partially
fixing a gauge choice for the PST gauge invariance.

The two-form field $B$ has the following contributions from K3 zero modes:
\begin{equation}
\label{2formdecomposition}
B_{ij}= \sum_{I=1}^{22} Y^I(\xi) b_{Iij}(x), \;\; B_{\alpha i}=0, 
\;\; B_{\alpha\beta}=c_{\alpha\beta}(\xi),
\end{equation}
where $b_{Iij}$ are the 22 harmonic representatives of H${}^2$(K3, {\bf Z}), the
integral second cohomology classes of K3. Any other terms are either massive or can be
removed by gauge transformations.
The nonzero components of $H_{\mu\nu\rho}$ and $\Htilde^{\mu\nu}$ are
\begin{equation}
H_{\alpha i j}=\sum_{I=1}^{22} \partial_\alpha Y^I b_{Iij} \label{Hform1}
\end{equation}
\begin{equation}
\Htilde^{i j}=\sum_{I=1}^{22}\Ytilde^I \frac{1}{2}\ep^{i j k l}b_{Ik l}=
\sum_{I=1}^{22}\sqrt{h} \Ytilde^I (\ast b_I)^{i j}, \label{Hform2}
\end{equation}
where $\Ytilde^I=\ep^{\alpha\beta}\partial_\alpha Y^I \partial_\beta a$ as in 
eq.~(\ref{Ytilde}). Note that $c_{\alpha\beta}$ does not contribute.

We need to understand why 
the coefficients $Y^I$ should be periodically identified with period $2\pi$,
since they will be identified with the 
$Y^I$ coordinates of the heterotic string action in sect. 2.
To show why that is so,
consider a transformation $B\rightarrow B+b^{(2)}$ with $db^{(2)}=0$. This
leaves invariant the lagrangian~(\ref{5brane}), which only depends on
$H=dB$. However, in the case of a general
background, the action also contains the term $-\frac{1}{2\pi}\int_{\cal Y} B\wedge G_4$,
which could fail to be invariant. 
Here $G_4 $ is the closed four-form background field strength of 11d supergravity.
It must satisfy the topological restriction~\cite{122}
\begin{equation}
\left[ \frac{G_4}{2\pi}\right] -\frac{p_1({\cal Y})}{4} \in H^4({\cal Y},{\bf Z}) ,
\end{equation}
where $p_1({\cal Y})$ is the first Pontryagin class of the 
M5-brane world-volume manifold ${\cal Y}$. 
In the present case, 
${\cal Y} = K3\times \Sigma$, 
where $\Sigma$ is identified as the 2d world-sheet of
the resulting heterotic string. 
The value of the first Pontryagin class is divisible by $4$
on all closed four-cycles of this manifold, and 
therefore the $p_1$ term can be dropped. 
It remains to determine the restrictions on $b^{(2)}$ that arise from requiring
that the change of the M5-brane action $\delta S=
-\frac{1}{2\pi}\int b^{(2)}\wedge G_4$ is an integral multiple of $2\pi$.
$b^{(2)}$ can be written
as a superposition of two-forms Poincar\'e
dual to the four-cycles of the 6d world volume. In the case at hand, the 
possible four-cycles are
K3 itself and $\Sigma$ times the 2-cycles of the K3.
The dual 2-forms are $\omega_2$, which is the normalized volume form of the world sheet $\Sigma$, and $b_I$, the 22 two-forms of K3 introduced earlier. So, writing
\begin{equation}
{1 \over 2 \pi} b^{(2)}=n^0\omega_2 +\sum_{I=1}^{22} n^I b_I, 
\end{equation}
where $n^0$ and $n^I$ are integers, it follows that
\begin{equation}
\delta S=- \left( n^0\int_{K3} G^{(4)}+\sum_{I=1}^{22} 
n^I \int_{C_I\times \Sigma} G^{(4)}\right)
\end{equation}
is an integer multiple of $2\pi$.
Thus, $B\rightarrow B + 2\pi (n^0\omega_2+\sum_{I=1}^{22}n^I b_I)$, 
which corresponds to 
shifting $Y^I$ by $2\pi n^I$, can be interpreted as a ``large gauge
transformation.'' This implies that the $Y^I$ are coordinates
of the Jacobian H${}^2$(K3, {\bf R})/H${}^2$(K3, {\bf Z}).
The $n^0$ term describes a shift of the coordinate
$c_{\alpha\beta}$ in eq.~(\ref{2formdecomposition}).
As noted earlier, it drops out of the formulas.

Now we can compute the string action that arises from double dimensional
reduction. Substituting the decompositions (\ref{Hform1}) and (\ref{Hform2})
into eq.~(\ref{5brane}), 
gives for the second term in the action
\begin{eqnarray}
 - \frac{\Htilde^{\mu\nu} H_{\mu\nu\rho} G^{\rho\lambda} \partial_\lambda a}
 {4 (\partial a)^2}&=&-\frac{\Ytilde^I (\ep^{ijkl}b_{Iij}b_{Jkl})
 \partial_\alpha Y^J \tilde G^{\alpha\beta}\partial_\beta a}{8 (\partial a)^2} \nonumber\\ &=&
-\frac{\Ytilde^I \sqrt{h}\ast(b_I\wedge \ast b_J)\partial_\alpha Y^J 
\tilde G^{\alpha\beta}\partial_\beta a}{2 (\partial a)^2}.
\end{eqnarray}
Here  $(\partial a)^2 = \tilde G^{\alpha\beta} \partial_\alpha a \partial_\beta a$.
Substituting eq.~(\ref{metricdecomposition}), 
the determinant in the first term of eq.~(\ref{5brane}) becomes
\begin{equation} \label{detterm}
\det\left( G_{\mu\nu}+i\frac{G_{\mu\rho}G_{\nu\sigma}\Htilde^{\rho\sigma}}
{\sqrt{-({\rm det}\,G_{\mu\nu}) (\partial a)^2}}\right)
= \tilde G h \left( 1+\frac{1}{2} \tr(\Atilde^2)+\det(\Atilde)\right),
\end{equation}
where $\tilde G = {\rm det}\, \tilde G_{\alpha\beta}$, 
$h = {\rm det}\, h_{ij}$, and 
\begin{equation}
\Atilde_i^{\; j}=\frac{1}{\sqrt{(-\tilde G)h(\partial a)^2}} h_{ik}\Htilde^{jk}.
\end{equation}
Substituting eq.~(\ref{2formdecomposition}) gives
\begin{equation}
\frac{1}{2}\tr(\Atilde^2)=\frac{\Ytilde^I b_{Iij} b_J^{ij} \Ytilde^J}
{2 \tilde G(\partial a)^2}=\frac{\Ytilde^I \ast(b_I\wedge \ast b_J) \Ytilde^J}
{\tilde G (\partial a)^2}
\end{equation}
and
\begin{eqnarray} \label{detA}
\det(\Atilde)&=&\det \left(\frac{h_{ij} \Ytilde^I (\ast b_{I})^{jk} }
{\sqrt{-\tilde G(\partial a)^2}} \right) =
\frac{1}{(-\tilde G(\partial a)^2)^2}\frac{1}{h} 
\left( \Pf\left( \Ytilde^I (\ast b_I)_{ij}\right)\right)^2 \nonumber\\ &=&
\frac{1}{4}\frac{\left( \Ytilde^I \ast(\ast b_I\wedge \ast b_J)\Ytilde^J 
\right)^2}{(-\tilde G(\partial a)^2)^2} .
\end{eqnarray}


To make the connection with the heterotic string action of the previous section,
we make the identifications 
\begin{equation}
 L_{IJ}=\int_{K3} b_I\wedge b_J,
\end{equation}
\begin{equation}
M_{IJ}=\int_{K3} b_I\wedge \ast b_J.
\end{equation}
Note that $\ast b_I = b_J (L^{-1} M)^J{}_I$, and therefore $(L^{-1}M)^2 =1$,
as in sect. 2.
Note also that $b_I\wedge b_J$ and $b_I\wedge \ast b_J$ are closed four-forms, and
therefore they are cohomologous to the unique harmonic four-form of the K3, 
which is the volume form $\omega$. It follows that\footnote{In the first version 
of this paper the exact terms were omitted from these formulas. We are grateful to G. Moore
for pointing out the error to us.}
\begin{equation} \label{bwedgeb}
b_I\wedge b_J=\ast b_I\wedge \ast b_J=\frac{L_{IJ}}{\cal V}\omega +dT_{IJ},\quad  
b_I\wedge \ast b_J=\frac{M_{IJ}}{\cal V}\omega +dU_{IJ},
\end{equation}
where ${\cal V}= \int_{K3} \omega$ is the volume of the K3
and $U_{IJ} =  T_{IK}(L^{-1} M)^K{}_J$. The exact terms are absent when either
two-form is self-dual, but there is no apparent reason why they should vanish when
both of them are anti-self-dual.
If we nevertheless ignore the exact pieces in these formulas,
substitute into  eq.~(\ref{5brane}) using eqs~(\ref{detterm}) -- (\ref{detA}),
and integrate over the K3, we obtain
\begin{equation}
\label{5brane on K3}
{\cal L}_1 = -{\cal V} \sqrt{-\tilde G}\sqrt{1+
\frac{\Ytilde^I M_{IJ} \Ytilde^J}{\tilde G(\partial a)^2 {\cal V}}
+\frac{1}{4}\left(\frac{\Ytilde^I L_{IJ}\Ytilde^J}
{\tilde G(\partial a)^2{\cal V}} \right)^2}
-\frac{\Ytilde^I L_{IJ} \partial_\alpha Y^J \tilde G^{\alpha\beta}\partial_\beta a}
{2(\partial a)^2}.
\end{equation}
This is precisely the heterotic string lagrangian (for $n=3$) presented in 
eq.~(\ref{hetfinal}) of the previous section provided that the 7d metric $g_{mn}$
in the string frame is related to the metric $\tilde g_{mn}$ derived from 11d by 
\begin{equation}
g_{mn} = {\cal V} \tilde g_{mn}
\end{equation}
so that $G_{\alpha\beta} = {\cal V}
\tilde G_{\alpha\beta}$. This is the same scaling rule found by a different 
argument in ref.~\cite{witten}.  Then, following ref.~\cite{witten},
the Einstein term in the 7d lagrangian
is proportional to ${\cal V} \sqrt{- \tilde g} R(\tilde g) = {\cal V}^{-3/2}
\sqrt{- g} R(g)$, from which
one infers that ${\cal V} \sim \lambda_H^{4/3}$.

To complete the argument we must still explain why terms that have been dropped
make negligible contributions. It is not at all obvious that 
the exact pieces in eq.~(\ref{bwedgeb}) can be neglected. However, this is what is required
to obtain the desired answer, so we are confident that it must be true. We hope to
return to this point in the future. The other class of terms that have been dropped are
the Kaluza--Klein excitations of the five-brane on the K3. By simple dimensional
analysis, one can show that in the heterotic string metric these contributions 
to the mass-squared of excitations are of order $\lambda_H^{-2}$. Therefore they
represent non-perturbative corrections from the heterotic viewpoint. Since our
purpose is only to reproduce the perturbative heterotic theory, they can be dropped.
Another class of contributions, which should not be dropped, correspond to
simultaneously wrapping the M2-brane around a 2-cycle of the K3. These wrappings
introduce charges for the 22 U(1)'s, according to how many times each cycle is wrapped.
The contribution to the mass-squared of excitations depends on the shape of the K3,
of course, but in the heterotic metric it is independent of its volume and hence of the
heterotic string coupling constant.

We are grateful to Michael Douglas and David Lowe for reading the manuscript.
We also wish to thank Greg Moore and Malcolm Perry for helpful discussions.

\end{document}